\documentclass[%
 reprint,
nofootinbib,
 amsmath,amssymb,
 aps,
 prl,
]{revtex4-2}

\usepackage{graphicx}
\usepackage{dcolumn}
\usepackage{bm}

\usepackage{lipsum}
\usepackage{amsmath}
\usepackage{graphicx}
\usepackage{color}
\usepackage{tikz}
\usepackage{verbatim}
\usepackage{hyperref}
\usepackage{amssymb}
\usepackage{bm}
\usepackage{subcaption}
\usepackage{graphicx}
\usepackage{subcaption}
\usepackage[export]{adjustbox}
\usepackage{wrapfig}

\usepackage{afterpage}  

\newcommand{\rom}[1]{\textup{\uppercase\expandafter{\romannumeral#1}}}

\begin{document}

\title{Testing the Starobinsky model of inflation with resonant cavities}
\author{Subhendra Mohanty}
    \email{mohantys@iitk.ac.in}

\affiliation{Indian Institute of Technology Kanpur, Kalyanpur, Kanpur, Uttar Pradesh - 208016, India}
\affiliation{Indian Institute of Science Education and Research Bhopal, Bhopal, Madhya Pradesh - 462066, India}

\author{Sukanta Panda}
     \email{sukanta@iiserb.ac.in}

\affiliation{Indian Institute of Science Education and Research Bhopal, Bhopal, Madhya Pradesh - 462066, India}

\author{Archit Vidyarthi}
    \email{architmedes@gmail.com}
       
\affiliation{Indian Institute of Science Education and Research Bhopal, Bhopal, Madhya Pradesh - 462066, India}
\affiliation{Indian Institute of Technology Bombay, Powai, Mumbai, Maharashtra - 400076, India}
\date{\today}

\begin{abstract}
We show that the Starobinsky inflation model based on $R^2$ gravity has a special feature that it provides a unique scalaron-two-graviton vertex with a coupling proportional to $1/M_P$. 
In this  model stochastic gravitational waves are produced when the scalaron - which is the massive scalar mode of the metric - decays into gravitons during reheating. This decay is accompanied by decay of scalaron into matter as well through a similar coupling, providing an efficient reheating. The stochastic gravitational waves thus produced have characteristic strain $h_c\sim 10^{-35}-10^{-34}$ in the frequency range $10^{6}-10^{12}\, {\rm Hz}$ which makes them accessible to resonant cavity searches for graviton to photon conversions. The detection of these high frequency gravitational waves would be a significant step in experimentally testing the Starobinsky inflation model.

\end{abstract}

\maketitle

\section{Introduction}
Inflationary models of the early universe \cite{Guth:1980zm,Albrecht:1982wi} provide a compelling framework for explaining the observed large-scale homogeneity, isotropy, and the generation of superhorizon primordial perturbations observed in the Cosmic Microwave Background (CMB) anisotropy \cite{Mather:1990tfx}. Among various models of inflation the Starobinsky model  \cite{Starobinsky:1980te} which is based on  a one parameter modifications $R^2$ to general relativity, is economical as the inflaton emerges from the scalar component of the spacetime metric.

The Starobinsky model is still viable \cite{Drees:2025ngb,Antoniadis:2025pfa,Zharov:2025zjg} in the light of experimental constraints on temperature and polarization anisotropies  from Planck \cite{Planck:2018jri} and the recent observations from ACT \cite{ACT:2025fju}. The model predicts a relatively small value for the tensor to scalar ratio $r$ whose range at $2\sigma$ is  $r=(2-3)\times 10^{-3} $ \cite{Drees:2025ngb} which makes the observation of tensor modes via the observation of B-mode polarisation in future space based experiments like COrE \cite{thecorecollaboration2011corecosmicoriginsexplorer}, AliCPT \cite{Li:2017drr}, LiteBIRD \cite{Matsumura:2013aja} and CMB-S4 \cite{Abazajian:2019eic},  challenging. A crucial aspect of such models is the post-inflationary reheating phase, during which the inflaton decays into standard model (SM) and beyond-standard model particles, potentially sourcing secondary gravitational wave (GW) production \cite{Allahverdi:2010xz}.

Production of GWs during the reheating era has garnered great interest since detectable GW signatures could help unveil post-inflationary physics shrouded behind the CMB. Existing literature in this regard relies heavily on graviton-matter couplings of the form $h_{\mu\nu}T^{\mu\nu}$ where observable GWs are produced either via inflaton scattering  \cite{Gross:2024wkl,Choi:2024ilx} or via graviton Bremsstrahlung during inflaton decay into matter \cite{Barman:2023ymn,Kanemura:2023pnv}. In some cases, novel couplings offering a scalar$\to$ two-graviton decay channel have also been introduced \cite{Ema:2021fdz,Tokareva:2023mrt} that dominate over the aforementioned processes and inevitably increase the number of parameters in the theory that need to be tuned using other observables.

In this work, we investigate the generation of a stochastic gravitational wave background (SGWB) through the decay of the inflaton in the Starobinsky model during reheating. A scalar $\to$ two-graviton decay channel has been derived previously via explicit perturbative expansion of the $R+R^2$ action in a flat background \cite{Hell:2023mph} (see Eq. (112))\footnote{Conclusions drawn in \cite{Hell:2023mph} regarding the vanishing degrees of freedom in the pure $R^2$ gravity have been shown to be erroneous in \cite{Golovnev:2023zen} when extended beyond backgrounds such as Minkowski which constitute a singular phase space submanifold. However, the perturbative expansion of the full $R+R^2$ action remains well-defined because the kinetic matrix remains non-degenerate.}. Here, we try to derive such decay vertices in the Jordan frame Starobinsky action and find that it can provide an efficient reheating stage as well as produce detectable gravitational waves. Unlike the primordial tensor perturbations typically studied in inflationary models, these secondary GWs can emerge at frequencies accessible to laboratory-scale experiments, specifically electromagnetic resonant cavities. We analyze the spectral properties of these GWs assuming no modifications to the standard Starobinsky inflation paradigm. By bridging the gap between inflationary dynamics and laboratory experiments, this study highlights the potential for direct detection of early-universe GW signals and exploring potential signatures of cosmology and  high-energy physics within accessible experimental frameworks.

\section{Scalaron Decay in Jordan Frame Starobinsky Model}
The action we consider for our analysis is,
\begin{equation}\label{eq:starobinsky-sm-action}
    S=\int d^4x \sqrt{-g}\left[\frac{M_P^2}{2}\left(R+\frac{\alpha}{2M_P^2}R^2\right)+\mathcal{L}_{SM}\right],
\end{equation}
where $\mathcal{L}_{SM}$ represents the Standard Model (SM) Lagrangian while the rest is the well-known Starobinsky inflation model. Focusing solely on the latter, one can replace the quadratic curvature term by an auxiliary scalar using the Lagrange multiplier method,
\begin{equation}\label{eq:starobinsky-jordan-lagrange}
    S=\int d^4x \sqrt{-g}\left[\frac{M_P^2}{2}\left(1+\frac{\alpha\chi^2}{M_P^2}\right)R-\frac{\alpha\chi^4}{4}\right].
\end{equation}
Though we have separated the longitudinal scalar mode, in the form of field $\chi$ and expressed the action in linear order of $R$, $\chi$ is still non-dynamical. The most common method to obtain the relevant kinetic term for $\chi$ is to perform a Weyl transformation followed by a field redefinition of $\chi$ to obtain a well-defined minimal scalar-tensor theory \cite{Panda:2022esd}. In the present work, we obtain a kinetic term for $\chi$ without sacrificing the non-minimal coupling. To do so, we first expand gravitational perturbations about a Minkowski background,
\begin{equation}
    g_{\mu\nu}=\eta_{\mu\nu}+\kappa h_{\mu\nu}.
\end{equation}
where $\kappa=2/ M_P$. This choice of background is relevant for high-energy processes and best suited for probes into deep horizon physics in expanding backgrounds. Now, we can write, up to quadratic order in $h_{\mu\nu}$,
\begin{align}\label{eq:quadratic-EH}
    &\sqrt{-g}\frac{M_P^2}{2}R=M_P(\partial_\mu\partial_\nu h^{\mu\nu}-\Box h)-\partial_\mu h^{\mu\nu}\partial_\nu h\nonumber\\&+\frac{1}{2}\partial_\mu h\partial^\mu h 
    -\frac{1}{2}\partial_\rho h^{\mu\nu}\partial^\rho h_{\mu\nu}+\partial_\mu h^{\mu\nu}\partial^\rho h_{\rho\nu}+\frac{1}{M_P}\mathcal{O}(h^3).
\end{align}
Substituting \eqref{eq:quadratic-EH} in \eqref{eq:starobinsky-jordan-lagrange}, we come across cross kinetic mixing terms of the form $\chi^2 \partial\partial h$ but no independent kinetic terms for the scalar degree of freedom (DOF) $\chi$. A kinetic term for $\chi$ can be obtained by performing the following linear translation,
\begin{equation}\label{eq:translation}
    h_{\mu\nu}\to h_{\mu\nu}-\eta_{\mu\nu}n\kappa\chi^2,
\end{equation}
where for $n=\alpha/4$ all cross kinetic terms cancel out. Note that this is simply a linearized Weyl transformation about a Minkowski background provided that $\chi/M_P\ll1$. Then, from \eqref{eq:starobinsky-jordan-lagrange}, we have,
\begin{align}
    S=&\int d^4x \left[\left(1+\frac{\alpha\chi^2}{M_P^2}\right)\left(-\partial_\mu h^{\mu\nu}\partial_\nu h+\frac{1}{2}\partial_\mu h\partial^\mu h \right.\right.\nonumber\\
    &\left.\left.-\frac{1}{2}\partial_\rho h^{\mu\nu}\partial^\rho h_{\mu\nu}+\partial_\mu h^{\mu\nu}\partial^\rho h_{\rho\nu}\right)-\frac{3\alpha^2}{4M_P^2}\partial_\mu\chi^2\partial^\mu\chi^2\right.\nonumber\\
    &\left.+\frac{\alpha^2}{M_P^3}\chi^2(\partial_\mu h^{\mu\nu}\partial_\nu\chi^2-\partial_\mu h\partial^\mu\chi^2)\right.\nonumber\\&\left.+\frac{3\alpha^3}{4M_P^4}\chi^2\partial_\mu\chi^2\partial^\mu\chi^2-\frac{\alpha\chi^4}{4}\right],
\end{align}
up to quadratic order in $h_{\mu\nu}$ (we have omitted relevant corrections from $\mathcal{O}(h^3)$ terms in \eqref{eq:quadratic-EH} for typographical ease). Having obtained the kinetic term for $\chi^2$, we proceed by canonicalizing the scalar DOF using the transformation,
\begin{equation}\label{eq:canonicalization-chi-zeta}
    \sqrt{\frac{3}{2}}\frac{\alpha}{M_P}\chi^2=\zeta,
\end{equation}
after which we arrive at the action,
\begin{align}\label{eq:quadratic-action-zeta}
    S&=\int d^4x \left[\left(1+\sqrt{\frac{2}{3}}\frac{\zeta}{M_P}\right)\left(-\partial_\mu h^{\mu\nu}\partial_\nu h+\frac{1}{2}\partial_\mu h\partial^\mu h\right.\right.\nonumber\\
    &\left.\left.-\frac{1}{2}\partial_\rho h^{\mu\nu}\partial^\rho h_{\mu\nu}+\partial_\mu h^{\mu\nu}\partial^\rho h_{\rho\nu}\right)-\frac{1}{2}\partial_\mu\zeta\partial^\mu\zeta-\frac{M_P^2}{6\alpha}\zeta^2\right.\nonumber\\
    &\left.+\frac{2}{3}\frac{\zeta}{M_P}(\partial_\mu h^{\mu\nu}\partial_\nu\zeta-\partial_\mu h\partial^\mu\zeta)+\frac{1}{\sqrt{6}}\frac{\zeta}{M_P}\partial_\mu\zeta\partial^\mu\zeta\right].
\end{align}
where $\zeta$ is identified as the scalaron which is the scalar inflaton associated with Starobinsky inflation with mass,
\begin{equation}
    M_\zeta\equiv M_P/\sqrt{3\alpha}\approx(3.173\pm0.022)\times10^{13}\text{GeV},
\end{equation}
based on Planck 2018 results \cite{Planck:2018jri} (see also \cite{DeFelice:2010aj,Ellis:2015pla}). It is straightforward to see at this stage that we have two distinct processes through which the inflaton could produce gravitons during the reheating stage: inflaton decay (suppressed by $M_P$), and inflaton annihilation (suppressed by $M_P^2$ because it requires two 3-vertices and off-shell mediation). Since inflaton decay is the leading order process, our analysis shall only consider that.

\section{Comparison with Einstein Frame Dynamics}
The $\zeta h h$ coupling in \eqref{eq:quadratic-action-zeta} is typically absent in the Einstein frame and has been shown to be absent even in $f(R)$ and Jordan frame $f(\phi)R$ theories in an FLRW spacetime background in \cite{Ema:2015dka,Ema:2016hlw}. However, it appears naturally here when we try to find canonical kinetic terms for the scalar DOF \cite{SupMat}. Crucially, in the Einstein frame, this linear coupling corresponds to the leading-order metric oscillation driven by the conformal factor $\Omega(\zeta)$ which relates the two frames. This effect is distinct from the inflaton annihilation effects discussed in \cite{Ema:2015dka}. Similar couplings between scalar and tensor modes has been derived explicitly in \cite{Hell:2023mph} by perturbatively expanding of the $R+R^2$ action (the expansion remains robust in light of \cite{Golovnev:2023zen}). See also \cite{Vilenkin:1985md} where the author finds $R$-dependent corrections to the graviton mode equations in Starobinsky gravity implying that the rapidly oscillating curvature $R$ (or equivalently $\zeta$) could act as a source for non-adiabatic production of gravitational waves during reheating.

The vanishing of the $\zeta\to hh$ amplitude in standard Jordan and Einstein frame descriptions relies on the definition of an asymptotic $\hat S$-matrix in a static vacuum where conformal invariance holds. However, our scenario is applied in  the reheating epoch, where the scalaron $\zeta(t)$ is a coherent, time-dependent background condensate. In this dynamical background, the notion of particle decay is replaced by particle production due to the breaking of time-translation invariance. The couplings derived in \eqref{eq:quadratic-action-zeta} act as time-dependent source terms for tensor modes describing the physical transfer of energy density from the oscillating inflaton to gravitational wave sector, which is non-zero regardless of the static vacuum definitions.

\section{Perturbative Reheating Epoch}
This phenomenon, where inflaton oscillations lead to particle production and reheating, is well studied in standard literature \cite{Kofman:1994rk,Kofman:1997yn,Ema:2015dka,Ema:2016hlw,Ema:2021fdz,Barman:2023ymn,Tokareva:2023mrt,Kanemura:2023pnv,Choi:2024ilx,Gross:2024wkl}. 
Based on the magnitude of decay width compared to the Hubble parameter, it can broadly be classified as broad resonance regime or preheating ($\Gamma\gg H$) and narrow resonance regime or perturbative reheating ($\Gamma\ll H$). As we shall see once we calculate the decay rates $\Gamma$ for various processes, we target perturbative reheating in this work, which can be safely described using the Feynman diagram approach for perturbative decay that we adopt.

It should also be noted that since in the standard picture \eqref{eq:starobinsky-sm-action}, we assume that $\mathcal{L}_{SM}$ is minimally coupled with gravity, 
\begin{equation}\label{eq:sm-minimal}
    \int d^4x \sqrt{-g}\mathcal{L}_{SM}=\int d^4x\left(1+\frac{h}{M_P}-\frac{4}{\sqrt{6}}\frac{\zeta}{M_P}\right)\mathcal{L}_{SM},
\end{equation}
due to \eqref{eq:translation} and \eqref{eq:canonicalization-chi-zeta}. We will impose a transverse-traceless (TT) gauge condition on the gravitons that will effectively remove the $h$-dependent couplings in the equation above. This implies that the minimal coupling of the SM Lagrangian with gravity provides vertices that would enable the scalaron to decay into SM particles. Given that the mass of SM particles $\ll M_\zeta$, the leading order terms contributing to the decay are the kinetic terms that resemble the structure,
\begin{equation}
    \frac{1}{\sqrt{6}}\frac{\zeta}{M_P}\partial_\mu X\partial^\mu X,
\end{equation}
where we have included contribution to the vertex from $g^{\mu\nu}$ present in the kinetic term, and $X$ represents some species of SM particles expected to be produced during reheating. Since the decay is through kinetic couplings, and $M_\zeta\gg m_X$, all SM particles thus produced are expected to be relativistic. Though different species of SM particles possess varied kinetic terms, this approximation using a set of scalar DOFs can provide a valuable estimate for whether this mechanism for reheating is sufficient. The squared matrix element for each $\zeta\to XX$ process is,
\begin{equation}
    |\mathcal{M}|^2=\frac{2}{3}\frac{M_\zeta^4}{M_P^2}.
\end{equation}
such that total decay width for the entire SM is,
\begin{equation}
    \Gamma_{SM}=g_{reh}\times\frac{M_\zeta^3}{24\pi M_P^2},
\end{equation}
where $g_{reh}\approx106.75$ is the effective number of relativistic DOFs in the SM plasma at reheating.  We make the standard assumption here that reheating ends as $H_{reh}\sim\Gamma$, where $\Gamma=\Gamma_{SM}+\Gamma_{hh}$ is the total rate of decay of the scalaron and $\Gamma_{hh}$ represents the rate of decay of scalaron to gravitons. Then using the expression for energy density at the end of reheating stage,
\begin{equation}
    \rho_{reh}=3M_P^2 H_{reh}^2=\frac{\pi^2}{30}g_{reh}T_{reh}^4,
\end{equation}
we can estimate the reheating temperature as,
\begin{equation}\label{eq:reheating-temp-general}
    T_{reh}=\left(\frac{90}{\pi^2}\frac{M_P^2\Gamma^2}{g_{reh}}\right)^{1/4},
\end{equation}
which can be calculated once we find $\Gamma_{hh}$. Note that thus far we have only considered direct decay of the inflaton into SM particles. This process may also be supplemented by graviton mediated diagrams and various other scattering process which are inevitably suppressed compared to the inflaton decay channels.

Observations place a constraint on the number of e-folds during reheating: $N_{reh}\lesssim20$ \cite{German:2022sjd}, where,
\begin{equation}
    N_{reh}\approx\frac{1}{3(1+\omega_{reh})}\ln{\left(\frac{\rho_{eoi}}{\rho_{reh}}\right)}.
\end{equation}
Here, $\omega_{reh}$ is the effective equation of state parameter during the reheating stage and $\rho_{eoi}=3M_P^2H_{eoi}^2$ is the energy density at the end of inflation. For \eqref{eq:quadratic-action-zeta}, we have,
\begin{equation}
    w_{reh}\approx\frac{\left<\Dot{\zeta}^2\right>-\left<M_\zeta^2\zeta^2\right>}{\left<\Dot{\zeta}^2\right>+\left<M_\zeta^2\zeta^2\right>}+\mathcal{O}\left(\frac{\left<\zeta\right>}{M_P}\right)\approx0.
\end{equation}
which corresponds to an early matter domination stage between inflation and reheating. This is because during reheating, $\zeta$ oscillates rapidly about $\zeta=0$, implying that $\left<\Dot{\zeta}^2\right>=\left<M_\zeta^2\zeta^2\right>$ and that kinetic interaction terms which scale as $\left<\zeta\right>/M_P=0$ vanish. Assuming the Hubble parameter at the end of inflation, $H_{eoi}\approx10^{13}$ GeV \cite{DeFelice:2010aj}, we find $N_{reh}\lesssim 15$ e-folds. The inequality signifies that as the scalaron decays into relativistic species, we approach radiation domination with $w\approx1/3$, thus further decreasing $N_{reh}$.

\section{Gravitational Waves: Production and Detection}
We can now safely move forward and study $\zeta\to hh$ decay and the subsequent production of SGWB. As mentioned earlier, the $\mathcal{O}(h^3)$ terms in \eqref{eq:quadratic-EH} can also provide $\zeta hh$ interaction terms once they undergo the shift \eqref{eq:translation} and canonicalization \eqref{eq:canonicalization-chi-zeta} operations. Imposing the TT gauge condition, we find that the only terms that contribute to the scalaron to two-graviton decay process are,
\begin{equation}\label{eq:scalar-decay-from-cubic}
    \sqrt{\frac{2}{3}}\frac{1}{M_P}(-\zeta h^{\mu\nu}\Box h_{\mu\nu}-2\zeta\partial_\rho h_{\mu\nu}\partial^\rho h^{\mu\nu}+\zeta\partial_\rho h^{\mu\sigma}\partial_\sigma h^\rho_\mu).
\end{equation}
Note that the first and last terms don't contribute because of the on-shell condition $k^2=0$ and transverse condition $k^\mu\epsilon_{\mu\nu}=0$, respectively. In momentum space representation, the corresponding decay amplitude is,
\begin{equation}
    i|\mathcal{M}|=\frac{4i}{M_P}\sqrt{\frac{2}{3}}\ |k_1\cdot k_2|\ (\epsilon_1\cdot\epsilon_2).
\end{equation}
where $k_{\{1,2\}\mu}$ represent the momenta of outgoing gravitons and $\epsilon_{\{1,2\}\mu\nu}$ represent the corresponding polarization tensors. In center of mass frame, for which $k_1\cdot k_2=-M_\zeta^2/2$, we find,
\begin{equation}
    \Gamma_{hh}=\frac{1}{6\pi}\frac{M_\zeta^3}{M_P^2}.
\end{equation}
Although derived as a single-particle decay width, $\Gamma_{hh}$ physically represents the dissipation rate of the zero-momentum inflaton condensate. The oscillating background $\zeta(t)$ serves as a coherent reservoir of mass $M_\zeta$, allowing the kinematic threshold for pair production to be met purely through inflaton oscillations. Using this, the number of extra relativistic DOFs produced during reheating can be calculated as well \cite{Garcia-Bellido:2012npk},
\begin{equation}
    \Delta N_{eff}=2.85\frac{\Gamma_{hh}}{\Gamma_{SM}}\approx 0.028,
\end{equation}
which is within the upper bound $\Delta N_{eff}\lesssim0.2$ posed by Planck constraints (see \cite{Gorbunov:2013dqa}). Also, now that we know $\Gamma_{hh}$ ($\ll \Gamma_{SM}$), we can calculate the reheating temperature substituting $\Gamma\approx\Gamma_{SM}$ in \eqref{eq:reheating-temp-general} resulting in,
\begin{equation}
    T_{reh}\approx(7.376\pm0.077)\times 10^{10}\text{GeV}.
\end{equation}
In instant reheating scenarios, inflaton decay starts right after the end of inflation and continues until reheating ends as $H_{reh}\sim \Gamma_{SM}$ as stated earlier. So, the infrared cutoff frequency can be found corresponding to the transition from inflation to reheating,
\begin{align}\label{eq:minimum-frequency-new}
    f_{min}&= \frac{M_\zeta}{4\pi}\frac{a_{eoi}}{a_0}=\frac{M_\zeta}{4\pi}\left(\frac{\rho_{reh}}{\rho_{eoi}}\right)^{1/3}\frac{T_0}{T_{reh}}\left(\frac{g_0}{g_{reh}}\right)^{1/3}\nonumber\\
    &\approx(3.400\pm0.035)\times10^{6}\text{Hz}
\end{align}
where $g_0=2+7/8\cdot(6\cdot4/11)$. Similarly, we can find the peak frequency can be found corresponding to the end of reheating \cite{Ema:2021fdz},
\begin{align}
    f_{peak}&=\frac{M_\zeta}{4\pi}\frac{a_{reh}}{a_0}=\frac{M_\zeta}{4\pi}\frac{T_0}{T_{reh}}\left(\frac{g_0}{g_{reh}}\right)^{1/3}\nonumber\\
    &\approx(4.061\pm0.014)\times10^{12}\text{Hz}.
\end{align}
The energy density spectrum for GWs produced via inflaton decay during an early matter domination stage \cite{Ema:2021fdz,Koshelev:2022wqj} \cite{SupMat} is given as,
\begin{equation}\label{eq:energy-density-spectrum}
    \frac{d\Omega_{hh}}{d\log k}=\frac{16k^4}{M_\zeta^4}\frac{\rho_{reh}}{\rho_0}\frac{\Gamma_{hh}}{H_{reh}}\frac{1}{\gamma(k)}e^{-\gamma(k)},
\end{equation}
where $k=2\pi f$, $\rho_0$ is the energy density of the universe at present time, and,
\begin{equation}\label{eq:gammak}
    \gamma(k)=\left[\left(\frac{g_{reh}}{g_0}\right)^{1/3}\frac{T_{reh}}{T_0}\frac{2k}{M_\zeta}\right]^{3/2}.
\end{equation}
SGWB spectra are typically expressed in terms of the characteristic strain $h_c(f)$,
\begin{equation}
    h_c(f)=\sqrt{\frac{3H_0^2}{f^2}\frac{d\Omega_{hh}}{d\log f}}.
\end{equation}
Now, using \eqref{eq:energy-density-spectrum} and \eqref{eq:gammak} and substituting values of various quantities, we arrive at the expression,
\begin{align}
   h_c(f)
    \approx& (8.417\pm0.037)\times10^{-31} \left(\frac{f}{\text{GeV}}\right)^{1/4} \nonumber\\
    &\times \exp{\left[(-2.290\pm0.012)\times10^{17}\left(\frac{f}{\text{GeV}}\right)^{3/2}\right]}.
\end{align}
Here, $h_c\propto f^{1/4}$ at low frequencies and decays exponentially at higher frequencies.
\begin{figure}
    \centering
    \includegraphics[width=\columnwidth]{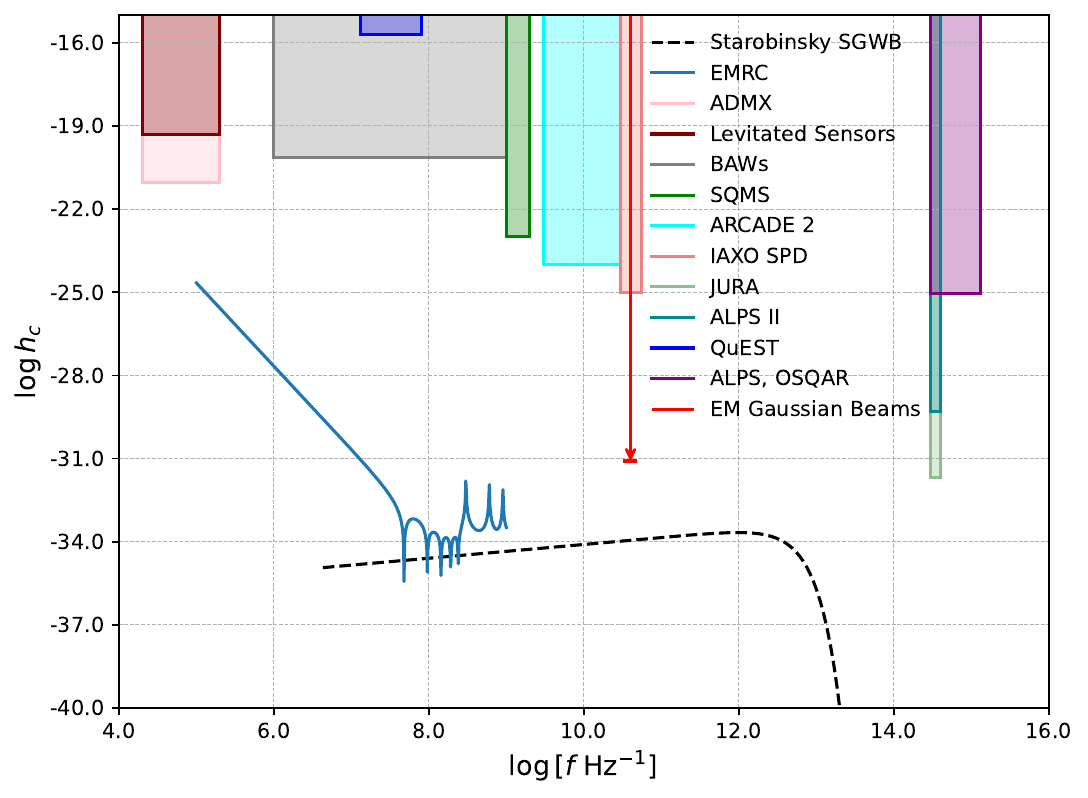}
    \caption{The SGWB signal produced during reheating in the Starobinsky model is shown to be detectable via proposed EMRC detectors. Detectability in EMRC is expected to be much higher in the downward spikes due to resonance effects \cite{Herman:2022fau}. The sensitivity curve for EMRC presented here corresponds to one set of model parameters that can be changed such that the detector functions in the kHz-THz range. Plots for other experiments are presented solely for reference and reflect approx. sensitivities.}
    \label{fig:sensitivity-plots}
\end{figure}
Though this combination of characteristic strain and frequencies is undetectable by current experiments, future GW detectors are expected to probe this range. In Fig. \ref{fig:sensitivity-plots}, we compare the detectability curves for existing and proposed high-frequency and ultra-high-frequency GW detectors \cite{ADMX:2023rsk,Goryachev:2014yra,Berlin:2021txa,Patra:2024eke} (see also \cite{Ejlli:2019bqj,Aggarwal:2025noe}) against the characteristic strain behavior obtained for our model. 

\section{Discussion}
In this paper, we have analyzed the spectrum of GWs produced during reheating in Starobinsky inflation due to a decay channel between the scalar DOF and gravitons present in the Jordan frame. This nonminimal coupling would have been lost due to the Weyl transformation when going to the Einstein frame \cite{Choi:2024ilx}. We also obtain a similar decay channel between scalaron and SM particles that allows efficient reheating resulting in the production of SM particles at $T_{reh}\approx(7.376\pm0.077)\times10^{10}$ GeV. The production of gravitons during the reheating stage results in a SGWB which falls in the high-frequency range of $\sim10^6-10^{12}$ Hz with a characteristic strain of $\sim10^{-35}-10^{-34}$. We have plotted characteristic strain $h_c(f)$ for the obtained SGWB spectrum against sensitivity curves of present and proposed GW detectors in Fig. \ref{fig:sensitivity-plots}. It is evident that the GW signature obtained in this work are within the proposed capabilities of electromagnetic resonant cavity (EMRC) experiments \cite{Herman:2022fau} and may even prove detectable by future iterations of Electromagnetic Gaussian Beam experiments \cite{Li:2003tv}. The sensitivity curve for EMRC shown in Fig. \ref{fig:sensitivity-plots} is specific to certain experimental parameters like cavity radius, length, and external magnetic field strength. In \cite{Herman:2022fau}, the authors discuss how changing the cavity radius can shift the sensitivity curves, potentially confirming the SGWB nature of detected GWs by placing the cut-off frequency between two resonant modes. In this letter, we provide  a proof of principle that Starobinsky inflation is accessible to probes with table-top experiments. Such a confirmation could validate the Starobinsky model of inflation effectively describing both an early universe inflation phase and an efficient reheating mechanism.

\begin{acknowledgments}
    \section{Acknowledgments}
    SP is partially supported by the DST (Govt. of India) Grant No. SERB/PHY/2021057. AV is supported by the Council of Scientific \& Industrial Research (CSIR), India under its Research Associateship program.
\end{acknowledgments}

\bibliographystyle{unsrtnat}
\bibliography{refs}

\end{document}


\title{Supplemental Material for ``Testing the Starobinsky model of inflation with resonant cavities"}

\author{Subhendra Mohanty}


\author{Sukanta Panda}


\author{Archit Vidyarthi}
       

\maketitle

\section{Inflaton decay in $f(\phi)R$ and $f(R)$ theories in Oscillating Backgrounds}

It was shown in [25,26] that the scalar$\to$2-graviton coupling responsible for inflaton decay into two graviton modes is typically absent in $f(R)$ and Jordan frame $f(\phi)R$ theories in an FLRW spacetime background when the scalar background is oscillating. The authors start with a Jordan frame action of the form,
\begin{equation}
    S=\int d^4x\sqrt{-g}\left[\frac{1}{2}f(\phi)R-\frac{1}{2}\partial_\mu\phi\partial^\mu\phi-V(\phi)\right]+S_M,
\end{equation}
where $\phi$ is the inflaton field, $S_M$ represents the matter part of the action, and,
\begin{equation}
    f(\phi)=M_P^2\left(1+c_1\frac{\phi}{M_P}+...\right)
\end{equation}
is the nonminimal coupling function. Next, from the Einstein equations,
\begin{equation}
    3fH^2=\rho_\phi-3H\dot f,
\end{equation}
they separate the oscillating and oscillation-averaged parts of various time-dependent quantities and obtain the relation,
\begin{equation}
    a(t)=a_0(t)\left(1-\frac{c_1}{2}\frac{\phi}{M_P}\right)
\end{equation}
up to first order in $\phi$ where $a_0(t)$ is independent of $\phi$. Now, considering inflaton decay into tensor modes, the authors consider the graviton action,
\begin{equation}
    S=\int d\tau d^3x a^2(t)f(\phi)\frac{1}{8}\left[h_{ij}^{'2}-(\partial_l h_{ij})^2\right],
\end{equation}
where $h_{ii}=\partial^i h_{ij}=0$. From the aforementioned analysis, it is clear that,
\begin{equation}
    a^2(t)f(\phi)\simeq M_P^2 a_0^2(t),
\end{equation}
implying that there is no decay of scalar into two graviton modes in $f(\phi)R$ models. A similar approach was used by the authors in [26] to show that there was no inflaton decay into two gravitons in $f(R)$ theories either. Before moving forward, we'd like to emphasize that the aforementioned analysis is valid in the era where the scalar-oscillations result in the modification of the scale factor. In our analysis, after replacing the $R^2$ term in Starobinsky model using an auxiliary scalar $\chi$, we arrive at the action,
\begin{equation}\label{eq:auxiliary-action}
    S=\int d^4x \sqrt{-g}\left[\frac{M_P^2}{2}\left(1+\frac{\alpha\chi^2}{M_P^2}\right)R-\frac{\alpha\chi^4}{4}\right],
\end{equation}
which possesses a nonminimal coupling but lacks a kinetic term for the scalar field $\chi$. At this stage, $\chi$ is not a propagating mode and, therefore, unable to produce any oscillations.

As mentioned in the main text, this action is simply an intermediate step between an $f(R)$ theory and its corresponding minimal scalar-tensor theory obtained using a Weyl transformation. We remain strictly within the Jordan frame background (Minkowski in the main text and flat FLRW in the supplementary file). Once the metric fluctuations are separated from the background metric, they are treated as independent degrees of freedom. We, instead, proceed by performing a linear translation of $h_{\mu\nu}$. This procedure has been used extensively in literature, especially in the context of modified gravity, to obtain distinct, quantizable modes while also avoiding any cross-kinetic terms. In our case, the linear translation helps extract a kinetic term for $\chi$ from $R$ without sacrificing the non-minimal coupling. The coupling between $\chi$ and two graviton modes obtained through this analysis has been cross-checked against the perturbative expansion of the $R+R^2$ action performed in [21].
    
Nonetheless, treating \eqref{eq:auxiliary-action} as an $f(\phi)R$ theory, we now verify if it also lacks a scalar to graviton decay channel as shown in [25]. Perturbing it around a general background,
\begin{equation}
    g_{\mu\nu}=\bar{g}_{\mu\nu}+\kappa h_{\mu\nu},
\end{equation}
where $\bar{g}$ is the background metric (which will later be fixed to be flat FLRW) and $\kappa=2/M_P$, we obtain up to first order in $h$,
\begin{align}
    &\sqrt{-g}\frac{M_P^2}{2}R=\sqrt{-\bar g}\left[\frac{M_P^2}{2}\bar R +M_P\left(h\bar R - 2h^{\mu\nu}\bar R_{\mu\nu}+\nabla_\mu\nabla_\nu h^{\mu\nu}-\Box h\right)+\mathcal{O}(h^2)+\frac{1}{M_P}\mathcal{O}(h^3)\right].
\end{align}
As in Minkowski background, we still need to perform a linear transformation of $h_{\mu\nu}$ to extract a kinetic term for $\chi$,
\begin{equation}
    h_{\mu\nu}\to h_{\mu\nu}-2n\bar g_{\mu\nu}\frac{\chi^2}{M_P}.
\end{equation}
Using symmetricity of $h_{\mu\nu}$ under the exchange of indices, we can fix $n=\alpha/4$ to cancel the cross-kinetic terms (same as for Minkowski background in the main text). Now, using the same canonicalization transformation,
\begin{equation}
    \sqrt{\frac{3}{2}}\frac{\alpha}{M_P}\chi^2=\zeta,
\end{equation}
we arrive at the following form of action:
\begin{equation}
    S=\int d^4x \sqrt{-\bar g} \left[\left(1+\sqrt{\frac{2}{3}}\frac{\zeta}{M_P}\right)\left\{\frac{M_P^2}{2}\bar R+M_P\left(h\bar R -\frac{\zeta}{\sqrt{6}}\bar R- 2h^{\mu\nu}\bar R_{\mu\nu}\right)-\frac{1}{2}h^{\mu\lambda}h^\nu_\lambda \bar R_{\mu\nu}\right\}+\mathcal{L}(h,\zeta)\right],
\end{equation}
where $\mathcal{L}(h,\zeta)$ represents the Lagrangian density of $h_{\mu\nu}$ and $\zeta$ modes that now behave as independent DOFs propagating in a general spacetime background with metric $\bar g_{\mu\nu}$. For $\bar g_{\mu\nu}=\eta_{\mu\nu}$,
\begin{align}
    \mathcal{L}(h,\zeta)=&\left[\left(1+\sqrt{\frac{2}{3}}\frac{\zeta}{M_P}\right)\left(-\partial_\mu h^{\mu\nu}\partial_\nu h+\frac{1}{2}\partial_\mu h\partial^\mu h-\frac{1}{2}\partial_\rho h^{\mu\nu}\partial^\rho h_{\mu\nu}+\partial_\mu h^{\mu\nu}\partial^\rho h_{\rho\nu}\right)-\frac{1}{2}\partial_\mu\zeta\partial^\mu\zeta-\frac{M_P^2}{6\alpha}\zeta^2\right.\nonumber\\
    &\left.+\frac{2}{3}\frac{\zeta}{M_P}(\partial_\mu h^{\mu\nu}\partial_\nu\zeta-\partial_\mu h\partial^\mu\zeta)+\frac{1}{\sqrt{6}}\frac{\zeta}{M_P}\partial_\mu\zeta\partial^\mu\zeta\right],
\end{align}
which is the Lagrangian density that we arrive at in the main text. The $\zeta h h$ coupling is clearly preserved here. In a general background, $\mathcal{L}(h,\zeta)$ gains additional curvature dependent contributions, including a curvature dependent mass term $-\bar R\zeta^2/12$, that become irrelevant for our computation in the main text.

Considering only the transverse-traceless modes of $h_{\mu\nu}$, the action becomes,
\begin{equation}
    S=\int d^4x \sqrt{-\bar g} \left[\left(1+\sqrt{\frac{2}{3}}\frac{\zeta}{M_P}\right)\left\{\frac{M_P^2}{2}\bar R\left(1-\sqrt{\frac{2}{3}}\frac{\zeta}{M_P}\right)-2M_Ph^{\mu\nu}\bar R_{\mu\nu}\right\}+\mathcal{L}(h,\zeta)\Big|_{TT}\right],
\end{equation}
where the subscript $TT$ symbolizes that the transverse-traceless condition has been applied to $\mathcal{L}(h,\zeta)$ as well. Recall that using the property of maximally symmetric spaces (which include flat FLRW), it can be shown for the spatial part of Ricci tensor that $\bar R_{ij}\propto \bar g_{ij}$, i.e. there exist no spatial components of transverse-traceless modes in $h^{\mu\nu}\bar R_{\mu\nu}$. Therefore, we have,
\begin{equation}
    S=\int d^4x \sqrt{-\bar g} \left[\left(1-\frac{2}{3}\frac{\zeta^2}{M_P^2}\right)\frac{M_P^2}{2}\bar R+\mathcal{L}(h,\zeta)\Big|_{TT}\right].
\end{equation}
The non-minimal coupling factor appears quadratic and not linear in $\zeta$. In fact, if we perform a full Weyl transformation instead of the linearized version used here, we would be left with no non-minimal coupling. Also, note that the propagating transverse traceless tensor modes are present in $\mathcal{L}(h,\zeta)\Big|_{TT}$ and not $\bar R$. The corresponding Einstein equations become,
\begin{align}
    \left(1-\frac{2}{3}\frac{\zeta^2}{M_P^2}\right)\left(\bar R_{\mu\nu}-\frac{1}{2}g_{\mu\nu}\bar R\right)=\frac{1}{M_P^2}T_{\mu\nu}(h,\zeta).
\end{align}
Now, assuming $\bar g_{\mu\nu}=(-1,a^2,a^2,a^2)$ where $a\equiv a(t)$ is the scale factor and following the same approach as [25,26], we find that the scale factor for the action (16) gets modified to,
\begin{equation}
    a(t)=a_0(t)\left(1+\frac{1}{3}\frac{\zeta^2}{M_P^2}\right),
\end{equation}
where $a_0(t)$ is the scale factor in the limit where the non-minimal coupling vanishes, i.e. $f(\phi)=1$. It is clear that the modified scale factor that is introduced via the measure function $\sqrt{-\bar g}$ doesn't affect the $\zeta hh$ interaction terms present in $\mathcal{L}(h,\zeta)$. It does affect the $\zeta\zeta hh$ interaction terms (and therefore scalaron annihilation processes), but those terms aren't the focus of the current work. To summarize, the $\zeta-$dependent modification of scale factor doesn't threaten the $\zeta\to hh$ decay process in our analysis. It can also be verified that the decay channel between scalaron $\zeta$ and SM particles also remains unchanged in the flat FLRW background.

\subsection*{Frame Transformation Invariance}
A natural follow-up would be whether this effect can be observed in the Einstein frame as well. In this section, we attempt to motivate how this effect can appear in the Einstein frame. While physical observables must stay invariant under frame transformations, invariance does not imply that the perturbative vertices must look identical in all variables. We know that even though the physical observables between conformally related frames remain unchanged, effective descriptions of the same physics can change drastically. The apparent vanishing of the decay vertex in the Einstein frame is not a disappearance of the physics, but a redistribution of the effect into the non-adiabatic evolution of the background. To demonstrate this, we note that the Weyl transformation of the metric to obtain Einstein frame action,
\begin{equation}
    g_{\mu\nu}\to\Omega^2(\zeta)g_{\mu\nu},
\end{equation}
involves a $\zeta$-dependent rescaling. During the reheating epoch in the Jordan frame, $\zeta$ oscillates about its vacuum expectation value, leading to reheating as shown in [28,29]. In the Einstein frame, the coupling doesn't exist explicitly, but the time oscillations of the background are encoded within the Einstein frame coordinates, captured by the integration measure $\sqrt{-g}$. The authors of [25] correctly show that there exists no explicit decay vertex $\zeta hh$ in the Einstein frame. Physically, however, the metric transformation implies that the spacetime coordinates in the Einstein frame gain oscillatory corrections proportional to $\Omega(\zeta)\approx 1+\mathcal{O}(\zeta)$. This leads to Bogoliubov production of transverse-traceless $h_{ij}$ modes due to non-adiabatic nature of the vacuum in the Einstein frame. Since this contribution is linear $\mathcal{O}(\zeta)$, it leads over the quadratic $\mathcal{O}(\zeta^2)$ coupling calculated in [25].

We find direct support for our argument in the analysis of the Starobinsky model in [27]. In Eq. (2.42) of [27], one can see that the friction term in the graviton mode equation gains corrections of the form $b'/b$ where $b=1+R/3M_\zeta^2$ and $'$ represents a derivative with respect to conformal time $d\eta=dt/a$. During the reheating epoch when $R$ is small and oscillating, we find,
\begin{equation}
    \frac{b'}{b}\approx\frac{R'}{3M_\zeta^2}-\frac{RR'}{9M_\zeta^4}.
\end{equation}
Here, the leading order correction is proportional to $R'$ ($\propto\zeta'$ in our analysis) and the subleading order correction is proportional to $RR'$ ($\propto\zeta\zeta'$). The latter is found as the leading order correction to the Hubble parameter due to oscillating background in Eq. (2.14) of [25].
    
Thus, our result is not a frame-dependent artifact, but the calculation of the energy transfer from the scalaron sector to the propagating graviton modes. This energy transfer is physically interpreted as the non-adiabatic production of gravitons in the oscillating background of the Einstein frame. The consistency of these two pictures is guaranteed by the frame invariance of the asymptotic particle number density, which our perturbative result correctly captures via the decay vertex in the narrow resonance limit.

\section{Derivation of Gravitational Wave Energy Density Spectrum}

The expression (27) in the main text is based on the work of [19] and stated in the context of an early matter domination era between the end of inflation and reheating in [33]. The authors in [19] consider production of gravitational waves (GWs) from the decay of dark Higgs particle. The energy density spectrum is defined as,
\begin{equation}
    \frac{d\Omega_{hh}}{d\ln k} \equiv \frac{1}{\rho_0}\frac{d\rho_h}{d\ln k},
\end{equation}
where $\rho_h$ is the energy density of gravitons and the other symbols represent the same quantities as in the main text. During the scalar oscillations (matter-like) stage, the scalaron energy density redshifts $\propto a^{-3}$ and decays with width $\Gamma\approx \Gamma_{SM}$. For a mode produced at redshift $z_d$ we may write,
\begin{equation}
    \rho_\phi(z_d)=\rho_{reh}\left(\frac{a_{reh}}{a_d}\right)^3 e^{-\Gamma_{SM} t_d},
\end{equation}
where $a_d$ and $t_d$ are the scale factor and cosmic time respectively at the start of decay. Using $a\propto t^{2/3}$ (matter-like) so $H\propto a^{-3/2}$, we have,
\begin{equation}
    \rho_\phi(z_d)=\rho_{reh}\,\left(\frac{H(z_d)}{H_{reh}}\right)^2 e^{-\Gamma_{SM} t_d}.
\end{equation}
Now, from [19] the energy spectrum of gravitons produced by the decay of a scalar $\phi$ of mass $m_\phi$ is,
\begin{equation}
    \frac{d\rho_h}{d\ln k}= \frac{16k^4}
    {m_\phi^4} \frac{\Gamma(\phi\to 2h)\,\rho_\phi(z_d)}{H(z_d)},
\end{equation}
where $1+z_d = m_\phi/(2k)$ implying that the gravitons observed today with energy $k$ were produced when the scalar had redshift $z_d$. Substituting and identifying $m_\phi$ as $M_\zeta$, we have,
\begin{align}
    \frac{d\rho_h}{d\ln k}&=\frac{16k^4}{M_\zeta^4}\Gamma(\zeta\to 2h)\frac{\rho_{reh}}{H_{reh}^2}\;H(z_d)e^{-\Gamma_{SM} t_d}\\
    &=\frac{16k^4}{M_\zeta^4}\;\rho_{reh}\;\frac{\Gamma(\zeta\to 2h)}{H_{reh}}\; \frac{e^{-\Gamma_{SM} t_d}}{\,H_{reh}\, / H(z_d)}.
\end{align}
We consider that the inflaton starts to decay right after the end of inflation at $H(z_d)$ and continues until $H_{reh}\sim\Gamma_{SM}$. Therefore, at the time of decay,
\begin{equation}
    t_d=\frac{1}{H(z_d)}\implies
\Gamma_{SM}t_d=\frac{\Gamma_{SM}}{H(z_d)}.
\end{equation}
Introducing the dimensionless ratio,
\begin{equation}
    \gamma\equiv\frac{H_{reh}}{H(z_d)},
\end{equation}
then,
\begin{equation}
    \Gamma_{SM}t_d=\frac{\Gamma_{SM}}{H(z_d)}
=\frac{\Gamma_{SM}}{H_{reh}}\gamma=\gamma,
\end{equation}
where we have used $H_{reh}\approx\Gamma_{SM}$ near the end of reheating. So,
\begin{equation}
    \frac{H(z_d)}{H_{reh}}e^{-\Gamma_{SM}t_d}=\frac{1}{\gamma}e^{-\gamma}.
\end{equation}
Substituting back, we find,
\begin{align}\label{eq:spectral-energy-density-derived}
    \frac{d\rho_h}{d\ln k} &= \frac{16k^4}{M_\zeta^4}\rho_{reh} \frac{\Gamma_{hh}}{H_{reh}} \frac{1}{\gamma}e^{-\gamma},\\
    \implies \frac{d\Omega_{hh}}{d\ln k} &= \frac{16k^4}{M_\zeta^4}\frac{\rho_{reh}}{\rho_0} \frac{\Gamma_{hh}}{H_{reh}} \frac{1}{\gamma}e^{-\gamma}.
\end{align}
To make $\gamma$ explicit in terms of observed energy $k$, we use the redshift representation during the matter-like scalar oscillations:
\begin{equation}
    \frac{H_{reh}}{H(z_d)}=\left(\frac{1+z_{reh}}{1+z_d}\right)^{3/2}.
\end{equation}
Now, using,
\begin{align}
    1+z_d&=\dfrac{M_\zeta}{2k},\\
    \text{and}\quad 1+z_{reh}&=\Big(\frac{g_{reh}}{g_0}\Big)^{1/3}\frac{T_{reh}}{T_0},
\end{align}
we get,
\begin{equation}\label{eq:gamma-derived}
    \gamma=\frac{H_{reh}}{H(z_d)} = \left[\left(\frac{g_{reh}}{g_0}\right)^{1/3}\frac{T_{reh}}{T_0}\,\frac{2k} {M_\zeta}\right]^{3/2}.
\end{equation}
Eq. \eqref{eq:spectral-energy-density-derived} and \eqref{eq:gamma-derived} represent the energy density spectrum of gravitational waves produced via inflaton decay during the scalar-oscillations, as used in [33].